\documentclass[12pt]{article}
\usepackage{graphicx}
\usepackage[cp1251]{inputenc}
\usepackage[russian]{babel}
\usepackage{rotating}
\begin{document}
 \noindent {\footnotesize\it Astronomy Letters, 2017, Vol. 43, No 7, pp. 452--463.}
 \newcommand{\dif}{\textrm{d}}

 \noindent
 \begin{tabular}{llllllllllllllllllllllllllllllllllllllllllllll}
 & & & & & & & & & & & & & & & & & & & & & & & & & & & & & & & & & & & & & \\\hline\hline
 \end{tabular}

  \vskip 0.5cm
 \centerline{\bf\large Vertical Distribution and Kinematics of Protoplanetary Nebulae}
 \centerline{\bf\large in the Galaxy}
 \bigskip
 \bigskip
  \centerline
 {
 V.V. Bobylev\footnote [1]{e-mail: vbobylev@gao.spb.ru} and
 A.T. Bajkova
 }
 \bigskip
 {
 \small\it
 Central (Pulkovo) Astronomical Observatory, Russian Academy of Sciences,

 Pulkovskoe sh. 65, St. Petersburg, 196140 Russia
 }
 \bigskip
 \bigskip
 \bigskip

 {
{\bf Abstract}---The catalogue of protoplanetary nebulae by
Vickers et al. has been supplemented with the line-of-sight
velocities and proper motions of their central stars from the
literature. Based on an exponential density distribution, we have
estimated the vertical scale height from objects with an age less
than 3 Gyr belonging to the Galactic thin disk (luminosities
higher than 5000 $L_\odot$) to be $h=146\pm15$~pc, while from a
sample of older objects (luminosities lower than 5000 $L_\odot$)
it is $h=568\pm42$~pc. We have produced a list of 147 nebulae in
which there are only the line-of-sight velocities for 55 nebulae,
only the proper motions for 25 nebulae, and both line-of-sight
velocities and proper motions for 67 nebulae. Based on this
kinematic sample, we have estimated the Galactic rotation
parameters and the residual velocity dispersions of protoplanetary
nebulae as a function of their age. We have established that there
is a good correlation between the kinematic properties of nebulae
and their separation in luminosity proposed by Vickers. Most of
the nebulae are shown to be involved in the Galactic rotation,
with the circular rotation velocity at the solar distance being
$V_0=227\pm23$ km s$^{-1}$. The following principal semiaxes of
the residual velocity dispersion ellipsoid have been found:
$(\sigma_1, \sigma_2, \sigma_3) = (47, 41, 29)$ km s$^{-1}$ from a
sample of young protoplanetary nebulae (with luminosities higher
than 5000 $L_\odot$), $(\sigma_1, \sigma_2, \sigma_3) = (50, 38,
28)$ km s$^{-1}$ from a sample of older protoplanetary nebulae
(with luminosities of 4000 $L_\odot$ or 3500 $L_\odot$), and
$(\sigma_1, \sigma_2, \sigma_3) = (91, 49, 36)$ km s$^{-1}$ from a
sample of halo nebulae (with luminosities of 1700 $L_\odot$).
  }

\medskip
DOI: 10.1134/S1063773717070027

 \subsection*{INTRODUCTION}
At present, the stars surrounded by gas--dust shells in a brief
transition phase from asymptotic giant branch (AGB) stars to
planetary nebulae are being actively studied. These are stars that
have already ceased to lose their mass on the AGB but have not yet
become hot enough to ionize the remnants of the shells surrounding
them. The universally accepted name of this evolutionary phase is
protoplanetary nebulae (PPNe) or post-asymptotic giant branch
(post-AGB) stars. However, the mass loss continues even after the
end of thermal pulsations in the post-AGB star. In particular,
white dwarfs, the central stars of planetary nebulae, lose their
mass with a rate of $\sim10^{-10} M_\odot$ yr$^{-1}$.

A large number of PPN candidates have been revealed (Garcia-Lario
et al. 1997) by infrared observations from the Infra-Red
Astronomical Satellite (IRAS) (Neugebauer et al. 1984). The
membership in the PPN class for most of them has been confirmed
relatively recently by analyzing their ground based optical
spectra (Garcia-Lario 2006; Su\'arez et al. 2006). The most
complete up-to-date information on PPNe in the Milky Way is
contained in the Toru\'n catalogue (Szczerba et al. 2007, 2012).

The shell expansion is believed to be spherically symmetric at the
AGB phase (Habing and Olofsson 2003). However, at the post-AGB
phase, apart from a more or less symmetric shell, collimated high
velocity bipolar jets are observed in the radio band
(P\'erez-S\'anchez et al. 2013). The environment in the shells and
jets is favorable for the emergence of maser emission (G\'omez et
al. 2015). Highly accurate measurements of the trigonometric
parallaxes have already been performed with ground-based very-long
baseline interferometers for some post-AGB objects (Imai et al.
2007, 2013).

In particular, stars with OH/IR maser emission (some of which are
post-AGB stars) play an important role in studying the structure
and kinematics of the Galaxy. For example, Debattista et al.
(2002) estimated the rotation velocity of the central bar using
radio observations of 250 OH/IR stars. In Gesicki et al. (2014)
post-AGB stars served to study the star formation history in the
Milky Way bulge.

Based on three density distributions and using data on planetary
nebulae from the catalogues by Frew (2008) and Stanghellini and
Haywood (2010), Bobylev and Bajkova (2017) found the vertical disk
scale height and determined the Galactic rotation parameters.
Obviously, estimating these parameters from a sample of objects at
an earlier evolutionary phase is of interest. At present, it has
become possible to perform such a study owing to the catalogue by
Vickers et al. (2015), which contains 209 PPNe and 87 candidates.
Quite reliable distance estimates have been obtained for all these
objects by fitting the observed spectral energy distribution to
the blackbody one. For so an extensive sample of PPNe such a
homogeneous distance scale has been realized by Vickers et al.
(2015) for the first time.

The goal of this paper is to study the $z$ distribution and
kinematics of PPNe in the Galaxy based on the catalogue by Vickers
et al. (2015). This suggests solving the following problems:
estimating the scale height, producing a sample with measured
line-of-sight velocities and proper motions, and estimating the
Galactic rotation parameters and the residual velocity dispersions
for PPNe as a function of their age or membership in such Galactic
subsystems as the thin disk, the tick disk, or the halo.

 \subsection*{METHODS}
In this paper we use a rectangular Galactic coordinate system with
the axes directed away from the Sun toward the Galactic center
($l=0^\circ, b=0^\circ,$ the $x$ axis or axis 1), in the direction
of Galactic rotation ($l=90^\circ, b=0^\circ,$ the $y$ axis or
axis 2), and toward the north Galactic pole ($b=90^\circ,$ the $z$
axis or axis 3).

We know three stellar velocity components from observations: the
line-of-sight velocity $V_r$ and the two tangential velocity
components $V_l=4.74r\mu_l\cos b$ and $V_b=4.74r\mu_b$ along the
Galactic longitude $l$ and latitude $b,$ respectively, expressed
in km s$^{-1}$. Here, the coefficient 4.74 is the ratio of the
number of kilometers in an astronomical unit to the number of
seconds in a tropical year, and $r$ is the stellar heliocentric
distance in kpc. The proper motion components $\mu_l\cos b$ and
$\mu_b$ are expressed in mas yr$^{-1}$. The velocities $U,V,W$
directed along the $x,y,z$ coordinate axes are calculated via the
components $V_r,V_l,V_b,$ respectively:
 \begin{equation}
 \begin{array}{lll}
 U=V_r\cos l\cos b-V_l\sin l-V_b\cos l\sin b,\\
 V=V_r\sin l\cos b+V_l\cos l-V_b\sin l\sin b,\\
 W=V_r\sin b                +V_b\cos b.
 \label{UVW}
 \end{array}
 \end{equation}
Note that when the residual velocity dispersions are calculated,
the velocities $U,V,W$ must be freed from the group motion of the
stars relative to the Sun ($U_\odot,V_\odot,W_\odot$) and from the
differential Galactic rotation effects.

 \subsubsection*{The Velocity Dispersion Ellipsoid}\label{sigm-123}
In this paper we use the following well-known method (Trumpler and
Weaver 1953; Ogorodnikov 1965) to estimate the residual velocity
dispersions for the PPN sample. We consider six second-order
moments $a, b, c, f, e,$ and $d:$
\begin{equation}
 \begin{array}{lll}
 a=\langle U^2\rangle-\langle U^2_\odot\rangle,\\
 b=\langle V^2\rangle-\langle V^2_\odot\rangle,\\
 c=\langle W^2\rangle-\langle W^2_\odot\rangle,\\
 f=\langle VW\rangle-\langle V_\odot W_\odot\rangle,\\
 e=\langle WU\rangle-\langle W_\odot U_\odot\rangle,\\
 d=\langle UV\rangle-\langle U_\odot V_\odot\rangle,
 \label{moments}
 \end{array}
 \end{equation}
which are the coefficients of the surface equation
 \begin{equation}
 ax^2+by^2+cz^2+2fyz+2ezx+2dxy=1,
 \end{equation}
and the components of the symmetric residual velocity moment
tensor
 \begin{equation}
 \left(\matrix {
  a& d & e\cr
  d& b & f\cr
  e& f & c\cr }\right).
 \label{ff-5}
 \end{equation}
The following six equations are used to determine the values of
this tensor from observations:
\begin{equation}
 \begin{array}{lll}
 V^2_r= a\cos^2 b\cos^2 l+b\cos^2 b\sin^2 l+c\sin^2 b\\
 +2f\cos b\sin b\sin l+2e\cos b\sin b\cos l+2d\sin l\cos l\cos^2 b,
 \label{EQsigm-1}
 \end{array}
 \end{equation}
\begin{equation}
 \begin{array}{lll}
 V^2_l= a\sin^2 l+b\cos^2 l\sin^2 l-2d\sin l\cos l,
 \label{EQsigm-2}
 \end{array}
 \end{equation}
\begin{equation}
 \begin{array}{lll}
 V^2_b= a\sin^2 b\cos^2 l+b\sin^2 b\sin^2 l+c\cos^2 b\\
 -2f\cos b\sin b\sin l-2e\cos b\sin b\cos l+2d\sin l\cos l\sin^2 b,
 \label{EQsigm-3}
 \end{array}
 \end{equation}
\begin{equation}
 \begin{array}{lll}
 V_lV_b= a\sin l\cos l\sin b+b\sin l\cos l\sin b\\
 +f\cos l\cos b-e\sin l\cos b+d(\sin^2 l\sin b-\cos^2\sin b),
 \label{EQsigm-4}
 \end{array}
 \end{equation}
\begin{equation}
 \begin{array}{lll}
 V_b V_r=-a\cos^2 l\cos b\sin b-b\sin^2 l\sin b\cos b+c\sin b\cos b\\
 +f(\cos^2 b\sin l-\sin l\sin^2 b)+e(\cos^2 b\cos l-\cos l\sin^2 b)\\
 -d(\cos l\sin l\sin b\cos b+\sin l\cos l\cos b\sin b),
 \label{EQsigm-5}
 \end{array}
 \end{equation}
\begin{equation}
 \begin{array}{lll}
 V_l V_r=-a\cos b\cos l\sin l+b\cos b\cos l\sin l\\
    +f\sin b\cos l-e\sin b\sin l+d(\cos b\cos^2 l-\cos b\sin^2 l),
 \label{EQsigm-6}
 \end{array}
 \end{equation}
which are solved by the least-squares method for the six unknowns
$a, b, c, f, e,$ and $d.$ The eigenvalues of the tensor (4)
$\sigma_{1,2,3}$ are then found from the solution of the secular
equation
 \begin{equation}
 \left|\matrix
 {
a-\lambda&          d&        e\cr
       d & b-\lambda &        f\cr
       e &          f&c-\lambda\cr
 }
 \right|=0.
 \label{ff-7}
 \end{equation}
The eigenvalues of this equation are known to be equal to the
reciprocals of the squares of the semiaxes of the velocity moment
ellipsoid and, at the same time, the squares of the semiaxes of
the residual velocity ellipsoid:
 \begin{equation}
 \begin{array}{lll}
 \lambda_1=\sigma^2_1, \lambda_2=\sigma^2_2, \lambda_3=\sigma^2_3,\\
 \lambda_1>\lambda_2>\lambda_3.
 \end{array}
 \end{equation}
The directions of the principal axes of the tensor (11),
$L_{1,2,3}$ and $B_{1,2,3}$ are found from the relations
 \begin{equation}
 \tg L_{1,2,3}={{ef-(c-\lambda)d}\over {(b-\lambda)(c-\lambda)-f^2}},
 \label{ff-41}
 \end{equation}
 \begin{equation}
 \tg B_{1,2,3}={{(b-\lambda)e-df}\over{f^2-(b-\lambda)(c-\lambda)}}\cos L_{1,2,3}.
 \label{ff-42}
 \end{equation}

 \subsubsection*{The Galactic Rotation Curve}
To determine the parameters of the Galactic rotation curve, we use
the equations derived from Bottlinger’s formulas in which the
angular velocity $\Omega$ is expanded into a series to terms of
the second order of smallness in $r/R_0:$
\begin{equation}
 \begin{array}{lll}
 V_r=-U_\odot\cos b\cos l-V_\odot\cos b\sin l-W_\odot\sin b\\
 +R_0(R-R_0)\sin l\cos b\Omega^\prime_0+0.5R_0(R-R_0)^2\sin l\cos b\Omega^{\prime\prime}_0,
 \label{EQ-1}
 \end{array}
 \end{equation}
 \begin{equation}
 \begin{array}{lll}
 V_l= U_\odot\sin l-V_\odot\cos l-r\Omega_0\cos b\\
 +(R-R_0)(R_0\cos l-r\cos b)\Omega^\prime_0
 +0.5(R-R_0)^2(R_0\cos l-r\cos b)\Omega^{\prime\prime}_0,
 \label{EQ-2}
 \end{array}
 \end{equation}
 \begin{equation}
 \begin{array}{lll}
 V_b=U_\odot\cos l\sin b + V_\odot\sin l \sin b-W_\odot\cos b\\
 -R_0(R-R_0)\sin l\sin b\Omega^\prime_0
    -0.5R_0(R-R_0)^2\sin l\sin b\Omega^{\prime\prime}_0,
 \label{EQ-3}
 \end{array}
 \end{equation}
where $\Omega_0$ is the angular velocity of Galactic rotation at
the Galactocentric distance $R_0$ of the Sun, the parameters
$\Omega'_0$ and $\Omega''_0$ are the first and second derivative
of the angular velocity, respectively, $V_0=|R_0\Omega_0|,$ and
$R$ is the distance from the star to the Galactic rotation axis
calculated from the formula
  \begin{equation}
 R^2=r^2\cos^2 b-2R_0 r\cos b\cos l+R^2_0.
  \end{equation}
We need to known the specific value of the distance $R_0.$
Gillessen et al. (2009) obtained one of its most reliable
estimates, $R_0=8.28\pm0.29$ kpc, by analyzing the orbits of stars
moving around the massive black hole at the Galactic center. The
following estimates were obtained from various samples of masers
with measured trigonometric parallaxes: $R_0=8.34\pm0.16$ (Reid et
al. 2014), $8.3\pm0.2$ (Bobylev and Bajkova 2014), $8.03\pm0.12$
(Bajkova and Bobylev 2015), and $8.24\pm0.12$ kpc (Rastorguev et
al. 2017). Based on these estimates, we adopted $R_0=8.3\pm0.2$
kpc in this paper.

 \subsection*{DATA}
 \subsubsection*{The Catalogue by Vickers et al. (2015)}
Basic information about 326 probable PPNe in the Milky Way and 107
candidates is collected in the Toru\'n catalogue (Szczerba et al.
2007, 2012), which is most complete to date. It contains data on
the classification, coordinates, and photometry of each object.

In this paper we use the distance estimates from the catalogue by
Vickers et al. (2015). It contains 209 PPNe and 87 candidates from
the Toru\'n catalogue. The distances to the objects were
determined by fitting the observed spectral energy distribution to
the blackbody one. The infrared photometric and spectroscopic data
from many sources (more than 20 catalogues) obtained in both
ground-based (for example, DENIS, 2MASS, UKIDSS) and spaceborne
(for example, IRAS, IUE, ISO, Planck, WISE) observations were used
for this purpose.

 \begin{table*}[p]
  \caption[]
   {\small
 Data on the PPNe
  }
  \begin{center}
  \label{t:DATA}   
  \footnotesize\baselineskip=0.1ex
  \begin{tabular}{|l|l|c|c|c|c|c|c|c|c|c|}\hline
        IRAS & Another     & $l$    & $b$     & Lum  &$r\pm\sigma_r$ & $\alpha (J2000.0)$ & $\delta (J2000.0)$ \\
             & designation & deg    &  deg    &      &  kpc          & h, m, s      & $^\circ$, m, s \\\hline
         1   &     2       &  3     &  4      &  5   &  6            &  7           &  8             \\\hline
             & LS 4825     &  1.671 &  -6.628 & 4000 & $6.45\pm1.36$ & 18 16 00.470 & -30 45 23.25 \\
 18371-3159  & LSE 63      &  2.918 & -11.818 & 1700 & $5.34\pm1.30$ & 18 40 22.017 & -31 56 48.82 \\
 17074-1845  & LSE 3       &  4.100 &  12.263 & 6000 & $6.88\pm1.18$ & 17 10 24.148 & -18 49 00.67 \\
 18384-2800  & V4728 Sgr   &  6.725 & -10.372 & 4000 & $2.29\pm0.62$ & 18 41 36.961 & -27 57 01.25 \\
 F16277-0724 & LS IV-07 1  &  7.956 &  26.706 & 3500 & $0.91\pm0.20$ & 16 30 30.018 & -07 30 52.03 \\
 17203-1534  & LS IV-15 3  &  8.548 &  11.486 & 6000 & $6.42\pm0.95$ & 17 23 11.916 & -15 37 15.06 \\
 17279-1119  & V340 Ser    & 13.230 &  12.174 & 6000 & $3.43\pm0.47$ & 17 30 46.925 & -11 22 08.28 \\
             & LS IV-04 1  & 14.403 &  22.856 & 1700 & $6.86\pm1.53$ & 16 56 27.730 & -04 47 23.70 \\
 F15240+1452 & BD+15 2862  & 21.866 &  51.930 & 1700 & $1.43\pm0.33$ & 15 26 20.818 & +14 41 36.32 \\
 19500-1709  & V5112 Sgr   & 23.984 & -21.036 & 6000 & $2.42\pm0.31$ & 19 52 52.701 & -17 01 50.30 \\
 \hline
 \end{tabular}
  \footnotesize\baselineskip=0.1ex
  \begin{tabular}{|c|c|c|c|l|r|c|c|l|c|}\hline
 $\mu_\alpha\cos\delta$~ & $\mu_\delta$   & $\sigma_{\mu_\alpha\cos\delta}$ & $\sigma_{\mu_\delta}$ & Source   &  $V_r$~~            & $\sigma_{V_r}$      & Source   & Spectral & ... \\
 {\tiny mas yr$^{-1}$}~  & {\tiny mas yr$^{-1}$} & {\tiny mas yr$^{-1}$}    & {\tiny mas yr$^{-1}$} & of $\mu$ & {\tiny km s$^{-1}$} & {\tiny km s$^{-1}$} & of $V_r$ &  type    & ... \\\hline
        9 ~~              &      10        &            11                   &            12         &    13    &  14~~               &   15                & 16       &  17      & 18--20 \\\hline
 $-2.7$ & $ -6.9$ & $3.0$ & $3.0$ & Tyc2 & $ -0.8$ &     & (1) & B1Ib         & ...\\ 
 $-5.2$ & $ -6.0$ & $3.5$ & $3.0$ & Tyc2 & $ 10.7$ & 1.0 & (2) & B1Iabe       & ...\\ 
 $-4.8$ & $ -9.2$ & $2.8$ & $2.6$ & Tyc2 & $ 12.4$ & 1.5 & (2) & B5Ibe        & ...\\ 
 $ 3.0$ & $ -5.1$ & $1.6$ & $1.5$ & Tyc2 & $-85.0$ & 5.0 & (3) & F2/3Ia+M?    & ...\\ 
 $ 3.6$ & $  1.6$ & $0.4$ & $0.3$ & Hip  & $  2.3$ & 2.0 & (4) & A9II/III     & ...\\ 
 $-6.8$ & $ -7.2$ & $2.2$ & $2.2$ & Tyc2 & $ 51.4$ & 2.1 & (2) & B1IIIpe      & ...\\ 
 $-2.5$ & $ -3.8$ & $1.5$ & $1.4$ & Tyc2 & $ 68.7$ &     & (5) & F2/3II       & ...\\ 
 $ 1.5$ & $ -4.9$ & $0.6$ & $0.5$ & Chen & $ 89.3$ &     & (6) & B            & ...\\ 
 $ 8.9$ & $-10.9$ & $0.5$ & $0.6$ & Hip  & $-45.0$ & 3.7 & (7) & B9Iab:p      & ...\\ 
 $-2.7$ & $ -3.8$ & $1.4$ & $1.4$ & Tyc2 & $ 13.0$ &     & (8) & F2/3(Iab+A)p & ...\\ 
 \hline
 \end{tabular}\end{center}  {\small
The PPN luminosity $Lum$ is given in solar luminosities,
Chen---Chen et al. (2000), (1)---Ryans (1997), (2)---Mello et al.
(2012), (3)---Reyniers and van Winckel (2001), (4)---Evans (1979),
(5)---Rao et al. (2012), (6)---Mooney et al. (2002),
(7)---Kharchenko et al. (2007), (8)---Klochkova (2013).
   }
 \end{table*}

 \subsubsection*{The Kinematic Sample}
When producing the kinematic sample to study the spatial
distribution and kinematics of PPNe in the Galaxy, we ran into the
problem that information about the line-of-sight velocities of
their central stars was presented quite poorly in the
SIMBAD~\footnote {http://simbad.u-strasbg.fr/simbad} electronic
database. At the same time, there are many publications where for
individual objects information about their systemic line-of-sight
velocities is available. These include, for example, the papers by
Klochkova et al. (1999, 2007, 2014, 2015), Klochkova (2013),
Arkhipova et al. (2001), or S\'anchez Contreras and Sahai (2012).
The bibliographic reviews with the line-of-sight velocities of
PPNe (Yoon et al. 2014) and OH masers (Deacon et al. 2004; Engels
and Bunzel 2016) turned out to be useful.

We took the proper motions from such catalogues as Hipparcos
(1997) revised by van Leeuwen (2007), TRC (Hog et al. 1998),
Tycho-2 (Hog et al. 2000), and the UCAC catalogues (Zacharias et
al. 2004, 2010, 2013).

Our list contains a total of 147 nebulae (we combined the lists of
PPNe and the lists of candidates from the catalogue by Vickers et
al. (2015)). It contains either only the line-of-sight velocities
(55 nebulae), or only the proper motions (25 nebulae), or both
line-of-sight velocities and proper motions (67 nebulae). The
first ten PPNe from our list are given in Table 1 (the remaining
ones are accessible in the electronic publication of the
catalogue):

Column 1 gives the IRAS number;

2--an alternative designation of the object;

3 and 4--the Galactic coordinates $l$ and $b$ copied from the
catalogue by Vickers et al. (2015);

5--the PPN luminosity $Lum$ (in solar luminosities) copied from
the catalogue by Vickers et al. (2015);

6--the heliocentric distance $r$ and its error $\sigma_r$ copied
from the catalogue by Vickers et al. (2015);

7 and 8--the equatorial coordinates $\alpha~(J2000.0)$ and
$\delta~(J2000.0),$ taken from the SIMBAD electronic database;

9 and 10--the proper motion components $\mu_\alpha\cos\delta$ and
$\mu_\delta$ taken from various sources;

11 and 12--the random measurement errors of the proper motion
components $\sigma_{\mu_\alpha\cos\delta}$ and
$\sigma_{\mu_\delta}$;

13--the designation of the catalogues of proper motions;

14 and 15--the heliocentric line-of-sight velocity $V_r$ and its
measurement error $\sigma_{V_r}$ (if it is available), below, for
example, when assigning the weights to the equations if there was
no estimate of the line-of-sight velocity error, we assumed
$\sigma_{V_r}=5$ km s$^{-1}$;

16--the reference to the line-of-sight velocity source;

17--the spectral type (if it is available)mainly from the SIMBAD
electronic database.

18, 19, and 20--the parallax and proper motion components (all
with the corresponding errors) from the Gaia DR1 catalogue (Prusti
et al. 2016).

 \subsection*{RESULTS AND DISCUSSION}
 \subsubsection*{The $z$ Distribution}
We analyzed the vertical distribution of PPNe using two samples.
To produce them, we divided all 296 objects into two approximately
equal parts by the luminosity specified in the catalogue by
Vickers et al. (2015). Sample 1 includes nebulae with a luminosity
higher than 5000 $L_\odot$, while sample 2 includes older nebulae
with a luminosity lower than 5000 $L_\odot$. The luminosities in
the catalogue by Vickers et al. (2015) have the following values:
20000 $L_\odot$, 15000 $L_\odot$, 12000 $L_\odot$, 6000 $L_\odot$,
4000 $L_\odot$, 3500 $L_\odot$ and 1700 $L_\odot$. According to
Table 2 from the paper of these authors, our sample 1 consists of
Galactic thin-disk objects with an age less than 3 Gyr, i.e., of
relatively young objects. Sample 2 has a more complex structure.
It can include old thin-disk objects, thick-disk and bulge
objects, and halo objects. In addition, we specified a constraint
on the heliocentric distances, which must not exceed 6 kpc. With
this condition we eliminate the objects subjected to the disk warp
in sample 1 and the bulge objects in sample 2.

To describe the observed frequency distribution of objects along
the $z$ coordinate axis, we apply the model of an exponential
density distribution:
 \begin{equation}
  N(z)=N_0 \exp \biggl(-{|z-z_\odot|\over h}\biggr),
 \label{exponent}
 \end{equation}
where $N_0$ is the normalization coefficient, $z_\odot$ is the
distance from the Sun to the Galactic midplane (the mean of the
$z$ coordinates of all objects from the sample), and $h$ is the
vertical disk scale height. The results of our analysis of samples
1 and 2 are reflected in the upper part of Table 2, where the
derived parameters of the exponential distribution (19) are given,
and in Fig. 1, where the histograms of the $z$ distributions of
PPNe are constructed. The results of our analysis of several
objects characterizing the properties of the Galactic disk are
presented in the lower part of Table 2.

As we can see from Table 2, $z_\odot$ is determined from PPNe with
very large errors. This parameter is determined much more
accurately from other classes of young Galactic thin-disk objects,
for example, hydrogen clouds, young open star clusters (OSCs), or
OB stars. For example, Bobylev and Bajkova (2016) calculated
$z_\odot=-16\pm2$ pc as a mean of the results obtained from a
number of young objects (OB associations, Wolf–Rayet stars, HII
regions, and Cepheids). Buckner and Froebrich (2014) found
$z_\odot=-18.5\pm1.2$ pc from their analysis of OSCs.

Note that based on clusters of various ages and using an
exponential density distribution, Buckner and Froebrich (2014)
traced the behavior of the scale height $h$ as a function of the
cluster age and position in the Galaxy. The value of $h=146\pm15$
pc we found is in agreement with $h=150\pm27$ pc obtained by
Bonatto et al. (2006) from open stars clusters with ages in the
range 200--1000 Myr or with $h=146\pm36$ pc found by Buckner and
Froebrich (2014) from clusters of the MWSC (Milky Way Star
Clusters) catalogue (Kharchenko et al. 2013) with a mean age of
$\sim$1 Gyr. Bobylev and Bajkova (2017) found $h=197\pm10$ pc from
a sample of planetary nebulae. Thus, a scale height $h\sim150$ pc
is typical for the mature population of the Galactic thin disk.

When the Galactic disk is arbitrarily divided into the thin and
thick ones, scale heights $h\sim150-250$ pc and $\sim$700 pc are
assumed to be typical for the thin and thick disks, respectively.
The view that there is no distinct thick disk, no bimodality, but
there is one evolving disk has been strengthened in recent years
(Bovy et al. 2012; Rix and Bovy 2013; Bovy et al. 2016). Bovy et
al. (2016) thoroughly studied a large sample of red clump giants
with various chemical compositions from the APOGEE (Apache Point
Observatory Galactic Evolution Experiment) survey (Majewski et al.
2016). The value of $h=568\pm42$ pc that we found from sample 2 is
intermediate between the extreme values, $h$ from 150 to 900 pc
(Bovy et al. 2016), and shows that this population has a mature
age and is fairly heated kinematically.

Note that the histogram constructed from old PPNe appears
asymmetric (Fig. 1b). This may be because the sample itself is
inhomogeneous. The scale heights separately from the southern
$(z<0$ pc) and northern ($z>0$ pc) parts of sample 2 are
$h=590\pm54$ pc at $z<0$~pc and $h=607\pm33$ pc at $z>0$~pc.

 \begin{table}[t]
 \caption[]
  {\small Parameters $z_\odot$ and $h$ }
  \begin{center}  \label{t:01}   
  \begin{tabular}{|l|c|c|c|c|c|}\hline
 Objects          & $N_\star$ & $z_\odot,$~pc & $h,$~pc    & Reference  \\\hline
 PPNe, sample 1   &   107     &  $-28\pm12$   & $146\pm15$ & this paper \\
 PPNe, sample 2   &   ~88     &  $-37\pm53$   & $568\pm42$ & this paper \\\hline
 OSCs             &   ---     & $-18.5\pm1.2$ &  ---       & (1) \\
 Various young objects  & --- &  $-16\pm2$ &  ---       & (2) \\
 OSCs, 200--1000 Myr    & --- &  $-15\pm2$    & $150\pm27$ & (3) \\
 Planetary nebulae      & 230 &  $-6\pm7$     & $197\pm10$ & (4) \\
 White dwarfs           &   717     &  ---          & 220--300   & (5) \\
 Old thin-disk stars    & $1\times10^4$ &  $-27\pm4$ & $330\pm3$ & (6) \\
 Oldest disk stars,     &               &            &           &     \\
 APOGEE, [Fe/H]$\sim-0.5$~dex & $1\times10^3$ &  ---       & $\sim800$ & (7) \\\hline
  \end{tabular}\end{center}  {\small
 $N_\star$ is the number of objects used,
 (1) Buckner and Froebrich (2014),
 2) Bobylev and Bajkova (2016), (3) Bonatto et al. (2006),
(4) Bobylev and Bajkova (2017), (5) Vennes et al. (2002), (6) Chen
et al. (2001), (7) Bovy et al. (2016).
 }
 \end{table}

 \begin{figure} {\begin{center}
 \includegraphics[width=150mm]{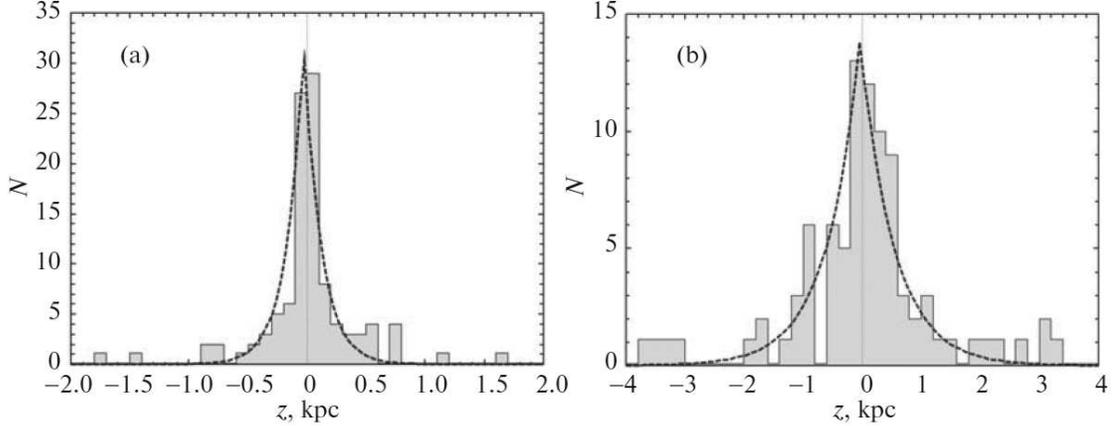}
 \caption{Histogram of the $z$ distribution of relatively younger
PPNe (a) and older PPNe (b). The dashed line represents the model
(19) constructed with the parameters from Table 2.
 }
 \label{f-hist} \end{center} } \end{figure}
 \begin{figure} {\begin{center}
 \includegraphics[width=130mm]{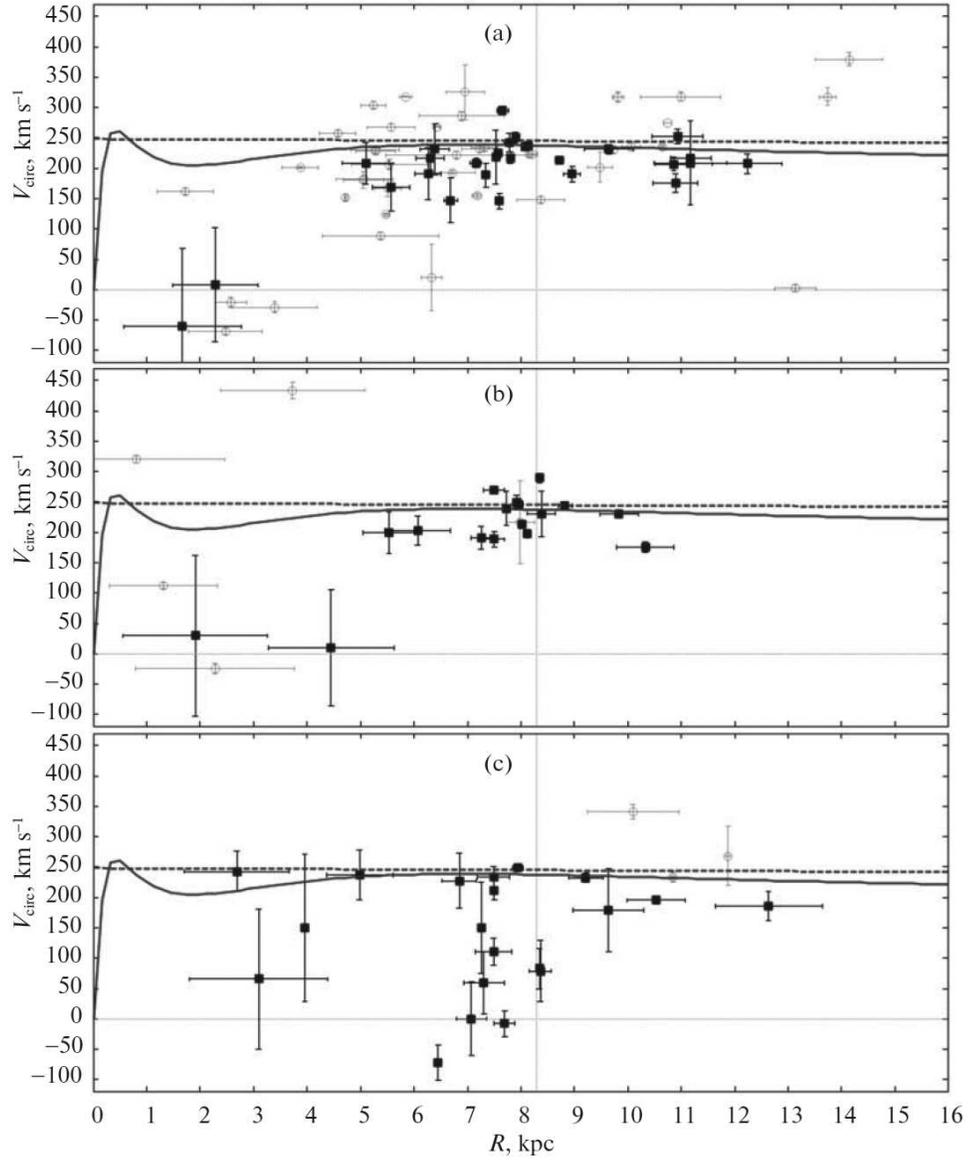}
 \caption{Estimates of the circular rotation velocities $V_{circ}$
for PPNe with luminosities higher than 5000 $L_\odot$ (a),
luminosities of 4000 $L_\odot$ and 3500 $L_\odot$ (b), and
luminosities of 1700 $L_\odot$ (c); those obtained from the total
space velocities (filled squares), those constructed from the
nebulae only with the line-of-sight velocities (open circles), the
Galactic rotation curve from Reid et al. (2014) is indicated by
the thick dashed line, the Galactic rotation curve from Bobylev
and Bajkova (2013) is indicated by the thick solid line, the
vertical dotted line marks the Sun’s position.
 }
 \end{center} } \end{figure}
 \begin{figure} {\begin{center}
 \includegraphics[width=80mm]{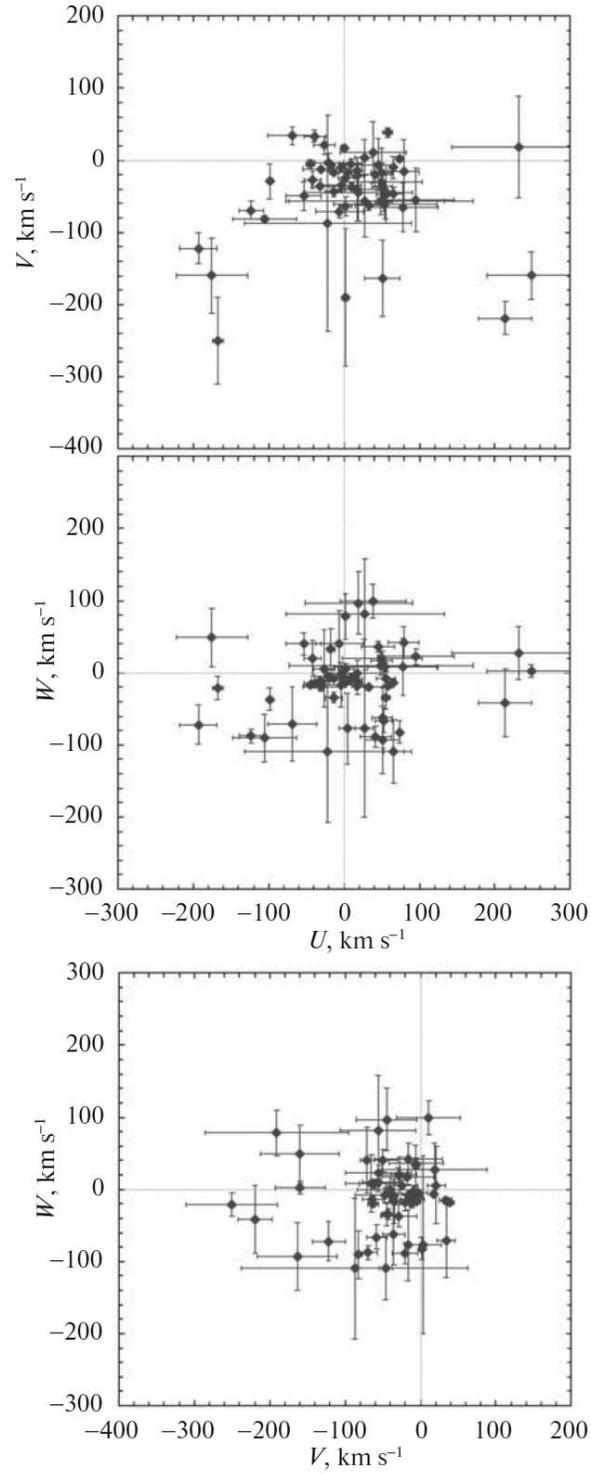}
 \caption{Residual velocities of PPNe on the $UV,$ $UW,$ and
$VW$ planes.
 }
 \end{center} } \end{figure}
 \begin{figure} {\begin{center}
 \includegraphics[width=150mm]{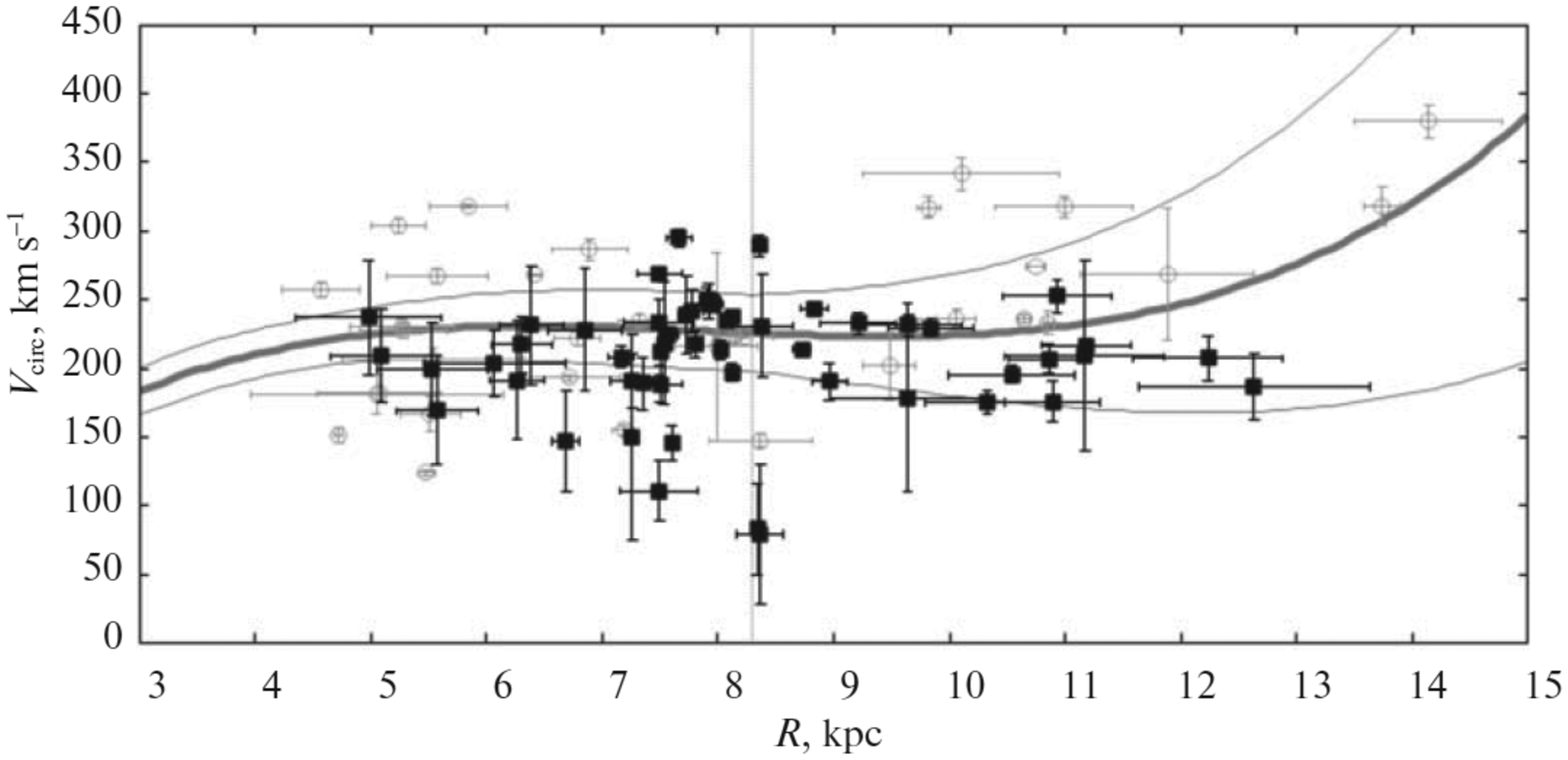}
 \caption{Circular rotation velocities $V_{circ}$ of PPNe constructed from their
total space velocities (filled squares); those constructed only
from the PPN line-of-sight velocities (open circles); the Galactic
rotation curve found in the solution (24) is indicated by the
thick line; the boundaries of the $1\sigma$ confidence intervals
are marked by the thin lines; the vertical dotted line marks the
Sun’s position.
 }
 \end{center} } \end{figure}

 \subsubsection*{Galactic Rotation}
In the case where all components of the space velocities are
known, the circular rotation velocity $V_{circ}$ of a nebula
around the Galactic rotation axis is found from the relation
 \begin{equation}
  V_{\rm circ}= U\sin \theta+(V_0+V)\cos \theta,
  \label{Vrot-UVW}
 \end{equation}
where $V_0=|R_0\Omega_0|$ and the position angle $\theta$ obeys
the relation $\tg\theta=y/(R_0-x)$. If, however, only the
line-of-sight velocity is known for a nebula, then $V_{circ}$ is
calculated from the formula
 \begin{equation}
  V_{\rm circ}=|R\Omega_0|+RV_r/(R_0\sin l\cos b).
  \label{Vrot-RV}
 \end{equation}
Figure 2 presents the circular rotation velocities $V_{circ}$ of
PPNe estimated from their total space velocities using Eq. (20)
with the adopted $V_0$ and the rotation velocities of other
nebulae constructed only from their line-of-sight velocities using
Eq. (21). The Galactic rotation curve obtained by Reid et al.
(2014) based on their analysis of $\sim$100 Galactic masers with
measured trigonometric parallaxes is also plotted here. This
rotation curve is described by the relation
$V_{circ}=V_0-0.2(R-R_0)$ km s$^{-1}$. It is clearly seen to be
close to a flat one with a constant rotation velocity $V_{circ}$.
Figure 2 shows the second, more complex Galactic rotation curve.
The parameters of this curve were found by Bobylev and Bajkova
(2013) by fitting the Allen--Santill\'an three-component model
gravitational Galactic potential (Allen and Santill\'an 1991) to
the data on masers with measured trigonometric parallaxes. In both
Reid et al. (2014) and Bobylev and Bajkova 2013) the rotation
velocity of the solar neighborhood $V_0$ is close to 240 km
s$^{-1}$ at $R_0=8.3$ kpc. More accurate values of $V_{circ}$ can
be obtained if $V_0$ is estimated more accurately from the
kinematics of PPNe themselves.

We considered three samples with different luminosity indices
$Lum.$ As can be seen from the figure, except for the bulge region
($R<4$ kpc), most of the PPNe are involved in the Galactic
rotation. There are quite a few objects with rapid rotation
($V_{circ}\sim220$ km s$^{-1}$) even in the sample with
luminosities of 1700 $L_\odot,$ which, according to Vickers et al.
(2015), are halo objects.

To study the kinematic properties of nebulae, we produced a
catalog of their residual velocities. It includes 67 PPNe for
which all three components of the space velocity are known. We
took the group velocity components
$(U_\odot,V_\odot,W_\odot)=(11.5,28.0,3.7)\pm(3.4,4.8,6.7)$~km
s$^{-1}$ found by Bobylev and Bajkova (2017) from their analysis
of planetary nebulae and took the flat curve from Reid et al.
(2014) as the Galactic rotation curve. The results are presented
in Fig. 3, where the derived residual space velocities of PPNe on
the $UV, UW,$ and $VW$ planes are shown. We can see from this
figure that there are two populations of points. The first (most
numerous) population is concentrated in a small central region
$(U\times V\times W)\sim(100\times 100\times 100)$~km s$^{-1}$,
while the second population is concentrated in a ring with a mean
radius of $\sim$200 km s$^{-1}$.

Based on Figs 2 and 3, we can conclude that: (i) when seeking the
Galactic rotation parameters based on PPNe, it is better not to
use any bulge objects ($R<4$ kpc), which barely rotate and have
large velocity errors; (ii) the main group of rotating objects has
residual velocities in a compact region with a radius
$\sqrt{U^2+V^2+W^2}$ less than 200 km s$^{-1}$; (iii) the rotating
group has velocity deviations $\Delta V_{\rm circ}$ no more than
150 km s$^{-1}$ with respect to the flat curve. All of this can be
expressed as the following constraints:
 \begin{equation}
 \begin{array}{lll}
 R>4~\hbox{kpc},\\
 \sqrt{U^2+V^2+W^2}<200~\hbox{km s$^{-1}$},\\
 \Delta V_{\rm circ}<150~\hbox{km s$^{-1}$}.
 \end{array}
 \end{equation}
Based on the PPNe with the proper motions and line-of-sight
velocities, we found the following kinematic parameters from the
solution of the conditional equations (15)--(17) using the
constraints (22):
 \begin{equation}
 \label{solution-1}
 \begin{array}{lll}
 (U_\odot,V_\odot,W_\odot)=(-9,29,11)\pm(6,8,6)~\hbox{km s$^{-1}$},\\
      \Omega_0=~28.3\pm3.1~\hbox{km s$^{-1}$ kpc$^{-1}$},\\
  \Omega^{'}_0=-3.40\pm0.63~\hbox{km s$^{-1}$ kpc$^{-2}$},\\
 \Omega^{''}_0=~0.266\pm0.410~\hbox{km s$^{-1}$ kpc$^{-3}$}.
 \end{array}
 \end{equation}
In this solution the error per unit weight is $\sigma_0=40.5$ km
s$^{-1}$, the Oort constants are $A=-14.1\pm2.6$ km s$^{-1}$
kpc$^{-1}$ and $B=14.2\pm 4.1$ km s$^{-1}$ kpc$^{-1}$. The
solution was found after two iterations with the elimination of
large residuals according to the $3\sigma$ criterion. We used a
total of 145 equations.

Another approach is that the nebulae with the proper motions and
line-of-sight velocities give all three equations (15)--(17), the
nebulae only with the line-of-sight velocities give one equation
(15), and the nebulae only with the measured proper motions give
two equations (16) and (17). Using the constraints (22), we found
 \begin{equation}
 \begin{array}{lll}
 (U_\odot,V_\odot,W_\odot)=(-1,30,10)\pm(5,6,5)~\hbox{km s$^{-1}$},\\
      \Omega_0=~27.8\pm2.1~\hbox{km s$^{-1}$ kpc$^{-1}$},\\
  \Omega^{'}_0=-3.67\pm0.46~\hbox{km s$^{-1}$ kpc$^{-2}$},\\
 \Omega^{''}_0=~0.934\pm0.266~\hbox{km s$^{-1}$ kpc$^{-3}$}.
 \end{array}
 \end{equation}
In this solution the error per unit weight is $\sigma_0=44.16$ km
s$^{-1}$, the Oort constants are $A=-15.2\pm1.9$ km s$^{-1}$
kpc$^{-1}$ and $B=12.5\pm2.9$ km s$^{-1}$ kpc$^{-1}$, and the
circular Galactic rotation velocity at the solar distance is
$V_0=231\pm23$ km s$^{-1}$. We used a total of 208 equations. The
solution was found after the elimination of large residuals
according to the $3\sigma$ criterion. As can be seen, in
comparison with the solution (23), the errors in the unknowns
decreased here.

There are a total of 57 stars with the proper motions from the
Gaia DR1 catalogue (Prusti et al. 2016) in our sample. This
catalogue was produced by combining the data in the first year of
Gaia observations with the Tycho-2 stellar positions and proper
motions (Hog et al. 2000). It is designated as TGAS (Tycho. Gaia
Astrometric Solution, Michalik et al. 2015; Brown et al. 2016;
Lindegren et al. 2016) and contains the parallaxes and proper
motions of $\sim$2 million bright stars. The random errors in the
parameters included in the Gaia DR1 catalogue are either
comparable to or smaller than those in the Hipparcos and Tycho-2
catalogues. The mean parallax errors are $\sim$0.3 mas. For most
of the TGAS stars the mean proper motion error is 1 mas yr$^{-1}$,
but for quite a few stars common to the Hipparcos catalogue this
error is smaller by an order of magnitude, $\sim$0.06 mas
yr$^{-1}$.

Here, we estimate the influence of allowance for the stellar
proper motions from the Gaia DR1 catalogue on the kinematic
parameters estimated from the entire sample of stars. For this
purpose, we use all of the available proper motions and
line-of-sight velocities and, where possible, the proper motions
from the Gaia DR1 catalogue.

By applying the approach described when seeking the solution (24)
using the constraints (22), we found the following parameters:
 \begin{equation}
 \label{solution-55}
 \begin{array}{lll}
 (U_\odot,V_\odot,W_\odot)= (0,24,10)\pm(5,6,5)~\hbox{km s$^{-1}$},\\
      \Omega_0=~27.3\pm2.1~\hbox{km s$^{-1}$ kpc$^{-1}$},\\
  \Omega^{'}_0=-3.69\pm0.46~\hbox{km s$^{-1}$ kpc$^{-2}$},\\
 \Omega^{''}_0=~0.611\pm0.264~\hbox{km s$^{-1}$ kpc$^{-3}$}.
 \end{array}
 \end{equation}
In this solution the error per unit weight is $\sigma_0=43.77$ km
s$^{-1}$, the Oort constants are $A=-15.3\pm1.9$ km s$^{-1}$
kpc$^{-1}$ and $B=12.0\pm2.8$ km s$^{-1}$ kpc$^{-1}$, and the
circular Galactic rotation velocity at the solar distance is
$V_0=227\pm23$ km s$^{-1}.$ The solution was found after the
elimination of large residuals according to the $3\sigma$
criterion. As can be seen, in comparison with the solution (24),
the error per unit weight $\sigma_0$ decreased here, while the
errors in the parameters being determined are the same as those in
the solution (24).

To produce the residual velocities, we will apply a flat curve
with $V_0=227$ km s$^{-1}.$ Figure 4 shows the sample of PPNe used
in seeking the solution (25) and the Galactic rotation curve found
from them. This curve is seen to be close to the flat one in the
range of distances $R$ 5--12 kpc.

Note that Bobylev and Bajkova (2017) determined the following
Galactic rotation parameters from 226 planetary nebulae at
$R_0=8.3$ kpc:  $(U,V,W)_\odot=(12,27,7)\pm(3,4,6)$~km s$^{-1}$,
      $\Omega_0=~27.4\pm2.6$~km s$^{-1}$ kpc$^{-1}$,
  $\Omega^{'}_0=-3.47\pm0.30$~km s$^{-1}$ kpc$^{-2}$ and
 $\Omega^{''}_0=~1.20\pm0.26$~km s$^{-1}$ kpc$^{-3}$, where the linear Galactic rotation
velocity at the solar distance is $V_0=227\pm30$ km s$^{-1}$. In
our view, in comparison with the solutions (24) and (25), the
components of the group velocity $U_\odot$ are determined better
in this solution.

The situation here is as follows. One reliable determination of
the peculiar solar motion relative to the local standard of rest
was made by Sch\"onrich et al. (2010),
$(U_\odot,V_\odot,W_\odot)=(11.1,12.2,7.3)\pm(0.7,0.5,0.4)$~km
s$^{-1}$. The lag behind the local standard of rest increases for
all older Galactic objects due to an asymmetric drift, i.e., the
velocity $V_\odot$ increases. The two other velocities $U_\odot$
and $W_\odot$ remain almost constant in a wide range of stellar
ages (Dehnen and Binney 1998). As a result, when producing the
residual velocities of PPNe, we suggest using the following
values:
 \begin{equation}
(U_\odot,V_\odot,W_\odot)=(11.1,29.7,7.3)~\hbox{km s$^{-1}$}.
 \label{UVW-sun}
 \end{equation}

 \subsubsection*{The Residual Velocities and Their Dispersions}
Based on the sample of relatively younger PPNe (with luminosities
higher than $5000 L_\odot$), we found the following residual
velocity dispersions calculated via the roots of the secular
equation (11):
 \begin{equation}
 \begin{array}{lll}
  \sigma_1=47\pm14~\hbox{km s$^{-1}$}, \\
  \sigma_2=41\pm 8~~\hbox{km s$^{-1}$}, \\
  \sigma_3=29\pm 3~~\hbox{km s$^{-1}$},
 \label{rezult-1}
 \end{array}
 \end{equation}
while the orientation of this ellipsoid is
 \begin{equation}
  \matrix {
  L_1=~53^\circ, & B_1=15^\circ, \cr
  L_2=145^\circ, & B_2=~7^\circ, \cr
  L_3=261^\circ, & B_3=74^\circ. \cr  }
 \label{rezult-11}
 \end{equation}
This solution was obtained using 222 equations in the system of
conditional equations (5)--(10).

Based on the sample of PPNe with luminosities of 4000 $L_\odot$
and 3500 $L_\odot$, we found the following residual velocity
dispersions:
 \begin{equation}
 \begin{array}{lll}
  \sigma_1=50\pm17~\hbox{km s$^{-1}$}, \\
  \sigma_2=38\pm22~\hbox{km s$^{-1}$}, \\
  \sigma_3=28\pm33~\hbox{km s$^{-1}$},
 \label{rezult-2}
 \end{array}
 \end{equation}
while the orientation of this ellipsoid is
 \begin{equation}
  \matrix {
  L_1=~72^\circ, & B_1=~11^\circ, \cr
  L_2=161^\circ, & B_2=-16^\circ, \cr
  L_3=345^\circ, & B_3=~71^\circ. \cr
   }
 \label{rezult-22}
 \end{equation}
This solution was obtained using 127 equations in the system of
conditional equations (5)--(10).

Finally, we analyzed the sample of PPNe with luminosities of 1700
$L_\odot$. According to the classification by Vickers et al.
(2015), some of them belong to the thick disk (with metallicities
[Fe/H] in the range from $-1.6$ to $-0.3$~dex), but they are
mostly Galactic halo objects. Based on this sample, we found the
following residual velocity dispersions:
 \begin{equation}
 \begin{array}{lll}
  \sigma_1=91\pm21~\hbox{km s$^{-1}$}, \\
  \sigma_2=49\pm25~\hbox{km s$^{-1}$}, \\
  \sigma_3=36\pm9~~\hbox{km s$^{-1}$},
 \label{rezult-3}
 \end{array}
 \end{equation}
while the orientation of this ellipsoid is
 \begin{equation}
  \matrix {
  L_1=~27^\circ, & B_1=-2^\circ, \cr
  L_2=117^\circ, & B_2=-6^\circ, \cr
  L_3=~96^\circ, & B_3=84^\circ. \cr
   }
 \label{rezult-33}
 \end{equation}
This solution was obtained using 92 equations in the system of
conditional equations (5)--(10).

The solutions (27)--(31) were obtained under identical
constraints. More specifically, the heliocentric distances of the
nebulae did not exceed 8 kpc, the bulge objects ($R>4$~kpc) were
eliminated, the magnitude of each of the velocities $U, V,$ or $W$
did not exceed 300 km s$^{-1}$ (see Fig. 3), and, finally, the
proper motions were used only up to heliocentric distances of 4
kpc (because the proper motion errors increase with distance).

It is interesting to compare our estimates with the results of the
analysis of white dwarfs from Pauli et al. (2006), where the
following velocity dispersions were determined:
 $(\sigma_U,\sigma_V,\sigma_W)=(34,24,18)$~km s$^{-1}$ for a sample of
361 thin-disk white dwarfs,
 $(\sigma_U,\sigma_V,\sigma_W)=(79,36,46)$ km s$^{-1}$
for a sample of 27 thick-disk white dwarfs, and
 $(\sigma_U,\sigma_V,\sigma_W)=(138,95,47)$ km s$^{-1}$ for a sample of 7 halo white dwarfs. A direct
comparison with the results that we obtained above is difficult to
make, because $(\sigma_U,\sigma_V,\sigma_W)$ are the residual
velocity dispersions directed along the $(x,y,z)$ coordinate axes,
while the ellipsoids (28)--(32) occasionally have significant
deviations from the directions of the $x$ and $y$ coordinate axes.

On the whole, we can conclude that there is a good correlation
between the spatial and kinematic properties of PPNe and their
separation in luminosity proposed by Vickers et al. (2015).

 \subsection*{CONCLUSIONS}
We supplemented the catalogue of PPNe by Vickers et al. (2015)
with a homogeneous distance scale by the line-of-sight velocities
and proper motions of their central stars from the literature.

An exponential density distribution was used to analyze the
vertical distribution of PPNe. We considered two samples of
objects from a solar neighborhood with a radius of 6 kpc separated
in age using the luminosity indices from the catalogue by Vickers
et al. (2015). Based on a sample of 107 objects with an age less
than 3 Gyr belonging to the Galactic thin disk, we estimated the
vertical scale height to be $h=146\pm15$ pc. Based on a sample of
88 older objects, we found $h=568\pm42$ pc, a value typical for
Galactic thick-disk objects.

We compiled a list of 147 PPNe in which there are only the
line-of-sight velocities for 55 nebulae, only the proper motions
for 25 nebulae, and both line-of-sight velocities and proper
motions for 67 nebulae. For a number of stars the proper motions
were taken from the Gaia DR1 catalogue. Based on this kinematic
sample, we obtained the following Galactic rotation parameters:
 $(U_\odot,V_\odot,W_\odot)=(0,24,10)\pm(5,6,5)$ km s$^{-1}$,
  $\Omega_0=~27.3\pm2.1$ km s$^{-1}$ kpc$^{-1}$,
  $\Omega^{'}_0=-3.69\pm0.46$ km s$^{-1}$ kpc$^{-2}$, and
  $\Omega^{''}_0=~0.611\pm0.264$ km s$^{-1}$ kpc$^{-3}$
for the adopted $R_0=8.3$ kpc. The Oort constants are
  $A=-15.3\pm1.9$~km s$^{-1}$ kpc$^{-1}$ and
  $B=12.0\pm2.8$ km s$^{-1}$ kpc$^{-1}$,
while the circular rotation velocity at the solar distance is
$V_0=227\pm23$ km s$^{-1}$.

We showed that there is a good correlation between the kinematic
properties of nebulae and their separation in luminosity proposed
by Vickers et al. (2015). For example, the following principal
semiaxes of the residual velocity dispersion ellipsoid have been
found:
 $(\sigma_1,\sigma_2,\sigma_3)=(47,41,29)$ km s$^{-1}$ from a sample of
relatively young PPNe (with luminosities higher than 5000
$L_\odot$),
 $(\sigma_1,\sigma_2,\sigma_3)=(50,38,28)$ km s$^{-1}$ from a sample of older
PPNe (with luminosities of 4000 $L_\odot$ or 3500 $L_\odot$), and,
finally,
 $(\sigma_1,\sigma_2,\sigma_3)=(91,49,36)$ km s$^{-1}$ from the oldest halo nebulae
(with luminosities of 1700 $L_\odot$).

 \subsubsection*{ACKNOWLEDGMENTS}
We are grateful to the referee for the helpful remarks that
contributed to an improvement of this paper. This work was
supported by the Basic Research Program P--7 of the Presidium of
the Russian Academy of Sciences, the ``Transitional and Explosive
Processes in Astrophysics'' Subprogram.

 \bigskip\medskip{REFERENCES}\medskip{\small

1. C. Allen and A. Santill\'an, Rev. Mex. Astron. Astrofis. 22,
255 (1991).

2. V. P. Arkhipova, N. P. Ikonnikova, R. I. Noskova, G. V.
Komissarova, V. G. Klochkova, and V. F. Esipov, Astron. Lett. 27,
719 (2001).

3. A. T. Bajkova and V. V. Bobylev, Baltic Astron. 24, 43 (2015).

4. V. V. Bobylev and A. T. Bajkova, Astron. Lett. 39, 809 (2013).

5. V. V. Bobylev and A. T. Bajkova, Astron. Lett. 40, 389 (2014).

6. V. V. Bobylev and A. T. Bajkova, Astron. Lett. 42, 1 (2016).

7. V. V. Bobylev and A. T. Bajkova, Astron. Lett. 43, (2017, in
press).

8. C. Bonatto, L. O. Kerber, E. Bica, and B. X. Santiago, Astron.
Astrophys. 446, 121 (2006).

9. J. Bovy, H.-W. Rix, and D. W. Hogg, Astrophys. J. 751, 131
(2012).

10. J. Bovy, H.-W. Rix, E. F. Schlafly, D. L. Nidever, J. A.
Holtzman, M. Shetrone, and T. C. Beers, Astrophys. J. 823, 30
(2016).

11. A. G. A. Brown, A. Vallenari, T. Prusti, J. de Bruijne, F.
Mignard, R. Drimmel, et al. (GAIA Collab.), Astron. Astrophys.
595, A2 (2016).

12. A. S. M. Buckner and D. Froebrich, Mon. Not. R. Astron. Soc.
444, 290 (2014).

13. L. Chen, M. Geffert, J. J. Wang, K. Reif, and J. M. Braun,
Astron. Astrophys. Suppl. Ser. 145, 223 (2000).

14. B. Chen, C. Stoughton, J. A. Smith, A. Uomoto, J. R. Pier, B.
Yanny, \v{Z}.E. Ivezi\'c, D. G. York, et al., Astrophys. J. 553,
184 (2001).

15. R. M. Deacon, J. M. Chapman, and A. J. Green, Astrophys. J.
Suppl. Ser. 155, 595 (2004).

16. V. P. Debattista, O. Gerhard, and M. N. Sevenster, Mon. Not.
R. Astron. Soc. 334, 355 (2002).

17. W. Dehnen and J. J. Binney, Mon. Not. R. Astron. Soc. 298, 387
(1998).

18. D. Engels and F. Bunzel, Astron.Astrophys. 582, A68 (2015).

19. D. S. Evans, in Proceedings of the IAU Symposium No. 30,
Toronto, Canada (1979), p. 57.

 20. D. J. Frew, PhD Thesis (Dep. Physics, Macquarie Univ., NSW 2109, Australia, 2008).

21. P. Garcia-Lario, A. Manchado, W. Pych, and S. R. Pottasch,
Astron. Astrophys. Suppl. Ser. 126, 479 (1997).

22. P. Garcia-Lario, in Planetary Nebulae in our Galaxy and
Beyond, Proceedings of the IAU Symposium No. 234, Ed. by M. J.
Barlow and R. H. Me\'ndez (2006).

23. K. Gesicki, A. A. Zijlstra, M. Hajduk, and C. Szyszka, Astron.
Astrophys. 566, A48 (2014).

24. S. Gillessen, F. Eisenhauer, T. K. Fritz, H. Bartko, K.
Dodds-Eden, O. Pfuhl, T. Ott, and R. Genzel, Astroph. J. 707, L114
(2009).

25. J. F. G\'omez, J. R. Rizzo, O. Su\'arez, A. Palau, L. F.
Miranda, M. A. Guerrero, G. Ramos-Larios, and J. M. Torrelles,
Astron. Astrophys. 578, 119 (2015).

26. H. J. Habing and H. Olofsson, Asymptotic Giant Branch Stars
(Springer, New York, Berlin, 2003).

27. The HIPPARCOS and Tycho Catalogues, ESA SP--1200 (1997).

28. E. Hog, A. Kuzmin, U. Bastian, C. Fabricius, K. Kuimov, L.
Lindegren, V. V. Makarov, and S. Roeser, Astron. Astrophys. 335,
L65 (1998).

29. E. Hog, C. Fabricius, V. V. Makarov, U. Bastian, P.
Schwekendiek, A. Wicenec, S. Urban, T. Corbin, and G. Wycoff,
Astron. Astrophys. 355, L27 (2000).

30. H. Imai, R. Sahai, and M. Morris, Astrophys. J. 669, 424
(2007).

31. H. Imai, T. Kurayama, M. Honma, and T. Miyaji, Astrophys. J.
771, 47 (2013).

32. N. V. Kharchenko, R.-D. Scholz, A. E. Piskunov, S. R\"oser,
and E. Schilbach, Astron. Nachr. 328, 889 (2007).

33. N. V. Kharchenko, A. E. Piskunov, E. Schilbach, S. R\"oser,
and R.-D. Scholz, Astron. Astrophys. 558, A53 (2013).

34. V. G. Klochkova, R. Szczerba, V. E. Panchuk, and K. Volk,
Astron. Astrophys. 345, 905 (1999).

35. V. G. Klochkova, E. L. Chentsov, N. S. Tavolzhanskaya, and V.
E. Panchuk, Astron. Rep. 51, 642 (2007).

36. V. G. Klochkova, Astron. Lett. 39, 765 (2013).

37. V. G. Klochkova, E. L. Chentsov, V. E. Panchuk, E. G.
Sendzikas, and M. V. Yushkin, Astrophys. Bull. 69, 439 (2014).

38. V. G. Klochkova, V. E. Panchuk, and N. S. Tavolzhanskaya,
Astron. Lett. 41, 14 (2015).

39. F. van Leeuwen, Astron. Astrophys. 474, 653 (2007).

40. L. Lindegren, U. Lammers, U. Bastian, J. Hernandez, S.
Klioner, D. Hobbs, A. Bombrun, D. Michalik, et al., Astron.
Astrophys. 595, A4 (2016).

41. S. R. Majewski, Astron. Nachr. 337, 863 (2016).

42. D. R. C. Mello, S. Daflon, C. B. Pereira, and I. Hubeny,
Astron. Astrophys. 543, 11 (2012).

43. D. Michalik, L. Lindegren, and D. Hobbs, Astron. Astrophys.
574, A115 (2015).

44. C. J. Mooney, W. R. J. Rolleston, F. P. Keenan, P. L. Dufton,
J. V. Smoker, R. S. I. Ryans, and L. H. Aller, Mon. Not. R.
Astron. Soc. 337, 851 (2002).

45. G. Neugebauer, B. T. Soifer, G. Miley, H. J. Habing, E. Young,
F. J. Low, C. A. Beichman, P. E. Clegg, et al., Astrophys. J.
278L, 83 (1984).

46. K. F. Ogorodnikov, Dynamics of Stellar Systems (Fizmatgiz,
Moscow, 1965) [in Russian].

47. E.-M. Pauli, R. Napiwotzki, U. Heber, M. Altmann, and M.
Odenkirchen, Astron. Astrophys. 447, 173 (2006).

 48. A. F. P\'erez-S\'anchez, W. H. T. Vlemmings, D. Tafoya, and J.
M. Chapman, Mon. Not. R. Astron. Soc. 436, L79 (2013).

 49. T. Prusti, J. H. J. de Bruijne, A. G. A. Brown, A. Vallenari, C.
Babusiaux, C. A. L. Bailer-Jones, U. Bastian, M. Biermann, et al.
(GAIA Collab.), Astron. Astrophys. 595, A1 (2016).

 50. S. S. Rao, S. Giridhar, and D. L. Lambert, Mon. Not. R.
Astron. Soc. 419, 1254 (2012).

 51. A. S. Rastorguev, M. V. Zabolotskikh, A. K. Dambis, N. D. Utkin,
     A. T. Bajkova, and V. V. Bobylev, Astrophys. Bull. {\bf 72}, 122 (2017).

 52. M. J. Reid, K. M. Menten, A. Brunthaler, X. W. Zheng,
T. M. Dame, Y. Xu, Y. Wu, B. Zhang, et al., Astrophys. J. 783, 130
(2014).

53. M. Reyniers and H. van Winckel, Astron. Astrophys. 365, 465
(2001).

54. H.-W. Rix and J. Bovy, Astron. Astrophys. Rev. 21, 61 (2013).

55. R. S. I. Ryans, P. L. Dufton, F. P. Keenan, S. J. Smartt, K.
R. Sembach, D. J. Lennon, and K. A. Venn, Astrophys. J. 490, 267
(1997).

56. C. S\'anchez Contreras and R. Sahai, Astrophys. J. Suppl. Ser.
203, 16 (2012).

57. R. Sch\"onrich, J. Binney, and W. Dehnen, Mon. Not. R. Astron.
Soc. 403, 1829 (2010).

58. L. Stanghellini and M. Haywood, Astrophys. J. 714, 1096
(2010).

59. O. Su\'arez, P. Garcia-Lario, A. Manchado, M. Manteiga, A.
Ulla, and S. R. Pottasch, Astron. Astrophys. 458, 173 (2006).

60. R. Szczerba, N. Siodmiak, G. Stasi\'nska, and J. Borkowski,
Astron. Astrophys. 469, 799 (2007).

61. R. Szczerba, N. Siodmiak, G. Stasi\'nska, J. Borkowski, P.
Garcia-Lario, O. Su\'arez, M. Hajduk, and D. A.
Garcia-Hern\'andez, in Planetary Nebulae: An Eye to the Future,
Proceedings of the IAU Symposium No 283, Ed. by A. Manchado, L.
Stanghellini and D. Sch\"onberner (2012), p. 506.

62. R. J. Trumpler and H. F. Weaver, Statistical Astronomy (Univ.
of California Press, Berkely, 1953).

63. S. Vennes, R. J. Smith, B. J. Boyle, S. M. Croom, A. Kawka, T.
Shanks, L. Miller, and N. Loaring, Mon. Not. R. Astron. Soc. 335,
673 (2002).

64. S. B. Vickers, D. J. Frew, Q. A. Parker, and I. S.
Boji\v{c}i\'c, Mon. Not. R. Astron. Soc. 447, 1673 (2015).

65. D.-H. Yoon, S.-H. Cho, J. Kim, Y. J. Yun, and Y.-S. Park,
Astrophys. J. Suppl. Ser. 211, 15 (2014).

66. N. Zacharias, S. E. Urban, M. I. Zacharias, G. L. Wycoff, D.
M. Hall, M. E. Germain, E. R. Holdenried, and L. Winter, Astron.
J. 127, 3043 (2004).

67. N. Zacharias, C. T. Finch, T. M. Girard, N. Hambly, G. Wycoff,
M. I. Zacharias, D. Castillo, T. T. Corbin, et al., Astron. J.
139, 2184 (2010).

68. N. Zacharias, C. T. Finch, T. M. Girard, A. Henden, J. L.
Bartlett, D. G. Monet, and M. I. Zacharias, Astron. J. 145, 44
(2013).

 }
\end{document}